\documentclass[11pt,twoside]{article}
\usepackage{asp2010}

\resetcounters

\newcommand{\Teff} {$T_{\rm eff}$}
\newcommand{\grav} {log\,{\em g}}
\newcommand{\vsini} {$v$\,sin\,$i$}
\newcommand{\macro} {$\Theta_{\rm RT}$}

\newcommand{\kms} {km\,s$^{-1}$}

\begin{document}

\title{Macroturbulent broadening: a single snap-shot approach to investigate pulsations in massive stars?}
\author{S. Sim\'on-D\'iaz$^{1,2}$, N. Castro$^3$, A. Herrero$^{1,2}$, C. Aerts$^{4,5}$, J. Puls$^6$, 
and N. Markova$^7$
\affil{$^1$Instituto de Astrof\'isica de Canarias, 38200 La Laguna, Tenerife, Spain.}
\affil{$^2$Departamento de Astrof\'isica, Universidad de La Laguna, 38205 La Laguna, Tenerife, Spain.}
\affil{$^3$Inst. of Astronomy \& Astrophysics, National Obs. of Athens, 15236 Athens, Greece.}
\affil{$^4$Instituut voor Sterrenkunde, Katholieke Universiteit Leuven, Celestijnenlaan 200D, 3001 Leuven, Belgium}
\affil{$^5$IMAPP, Department of Astrophysics, Radboud University Nijmegen, PO Box 9010, 6500 GL Nijmegen, the Netherlands}
\affil{$^6$Universit\"atssternwarte M\"unchen, Scheinerstr. 1, 81679 M\"unchen, Germany}
\affil{$^7$Institute of Astronomy with  NAO, BAS, P.O. Box 136, 4700 Smolyan, Bulgaria}
}

\begin{abstract}
We present a brief progress report of our project aimed at the investigation of the so-called 
{\em macroturbulent} broadening affecting the line-profiles of O and early B-type stars, and
speculate about the possible use of this spectroscopic feature as a single snap-shot approach 
to investigate pulsations in massive stars.
\end{abstract}

\section{Introduction}
In addition to the mass and the stellar wind properties, stellar rotation is also a crucial 
parameter for the evolution of massive stars \citep[e.g.][and the associated series of papers in the last two decades]{Maeder00}. 
The advent of CCD detectors and high resolution spectrographs has made it possible to confirm the 
hypothesis proposed\footnote{At this time, this hypothesis was based on indirect arguments such as, 
for example, the small number of narrow lined O type stars and B supergiants (B Sgs) found in the
analyzed samples.} by several authors \citep[e.g.][and references therein]{Sle56, Con77, How97} 
that rotation is not the only dominant broadening agent shaping the metal line-profiles in OB type 
stars. A type of extra-broadening, usually claimed to be associated with large scale velocity 
fields, and even called {\em macroturbulent} broadening at some point, is clearly contributing
to the total broadening in these stars \citep[][see also Sect. \ref{sec21} below]{Sim07,Duf06, 
Lef07, Mar08, Fra10, Sim11a, Mar11}.

The main consequence of this confirmation is that previous determinations of projected rotational
velocities in OB stars need to be revised. In addition, other important issues such as, e.g., what is 
the physical origin of this apparently ubiquitous line-broadening, or what is the impact of this 
phenomenon on our knowledge of massive star evolution also arise from this observational result.  
These questions will certainly be treated in the next
years. We refer the reader to \cite{Sim11} for a detailed review on the advances on this 
topic in the last years (some of them updated here). In this contribution we concentrate on recent advances of the
{\em IACOB} project (P.I. Sim\'on-D\'iaz) concerning the characterization of the {\em macroturbulent} broadening, and
speculate about one possible application of this spectroscopic feature that came up while working 
on this project: \cite{Aer09} have recently revived\footnote{Already \cite{Luc76}
suggested a pulsational origin of the {\em macroturbulent} broadening.} the 
suggestion that {\em macroturbulent} broadening may be identified with the surface motions 
generated by the superposition of numerous high order non-radial oscillations. If this hypothesis
is finally confirmed on observational grounds, could hence {\em macroturbulent} broadening
be used as a single snap-shot approach to investigate pulsations in massive stars?

\section{The IACOB project: line-broadening in OB stars}\label{sec21}

The {\em IACOB spectroscopic survey of Galactic OB stars} \citep{Sim11a, Sim11b} 
is an ambitious observational project which has compiled the largest homogeneous, high-resolution database of optical 
spectra of massive stars observable from the Northern hemisphere to date. The {\em IACOB} project aims at the 
scientific exploitation of this unique spectroscopic dataset. One of the drivers of this survey is the 
characterization of the {\em macroturbulent} broadening in the whole massive star domain, and the
investigation of its physical origin. Two type of observations were hence planned
within the project. On the one hand, we observed a large sample of O and early-B type stars 
(including dwarfs, giants and supergiants up to B2) to measure their 
projected rotational velocities and quantify the amount of extra-broadening present in these stars. 
On the other hand, we are obtaining time\,series of spectroscopic observations for a small
sample of selected (bright) candidates to investigate the possible pulsational origin of this extra-broadening.
Some results of this investigation are indicated below.

\subsection{Characterization of the line broadening in OB stars}\label{sec21}

\begin{figure}[t!]
\center
\includegraphics[height=12.0cm, angle=90]{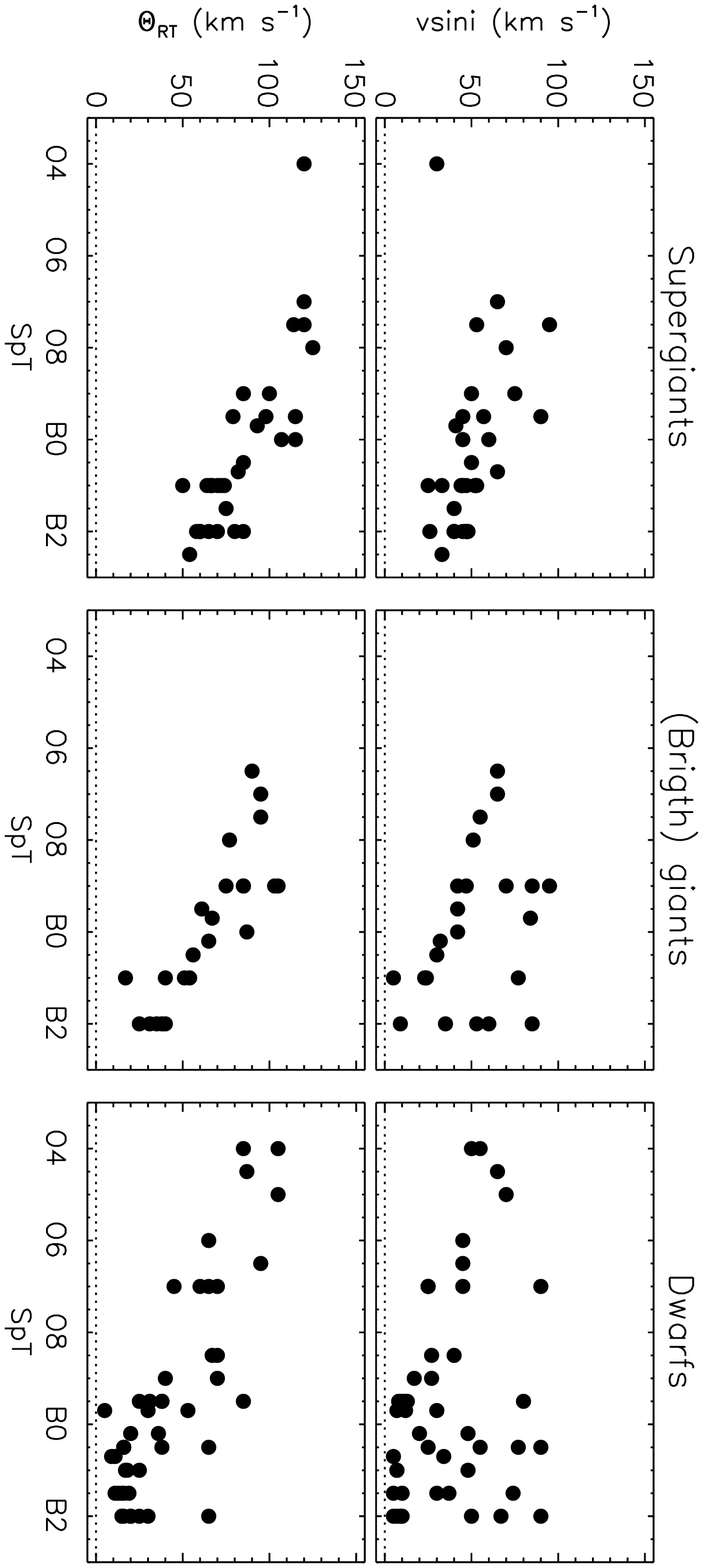}
\caption{Projected rotational velocities (\vsini) and size of the {\em macroturbulent} 
broadening \citep[characterized here assuming a radial-tangential Gaussian broadening profile, \macro,
see][]{Sim11} from a sample of {\em IACOB} spectra. Note that only stars with \vsini\,$\le$\,100 \kms\ are included 
in these plots. The accurate determination of the extra-broadening contribution
in stars with larger \vsini\ is complicated by subtle effects (e.g the placement 
of the continuum in the fitting process) which require a more careful study, and hence have been 
excluded for the moment.
\label{fig1}}
\end{figure}

The analysis of the {\em IACOB} spectra using a combined technique based on the Fourier 
transform (FT) and goodness-of-fit (GOF) methods (Sim\'on-D\'iaz et~al., in prep.) is allowing to 
review previous determinations of projected rotational velocities and characterize the extra-broadening
in $\sim$\,150 Galactic OB stars. Preliminary results from
this study (Fig. \ref{fig1}) indicate that the extra-broadening is a common feature not only in B Sgs, 
but also in B giants and O-type stars of all luminosity classes\footnote{A similar result 
has also been found from an independent study by \cite{Mar11}.}. 

The first obvious consequence of this analysis is that previous \vsini\ determinations in OB stars
not accounting for the extra-broadening must be revised downwards in many cases. In addition, the
ubiquitous presence of the {\em macroturbulent} broadening makes it very important to investigate the physical driver 
of this extra-broadening and its implications on the evolution and the
stellar and wind properties of massive stars.

\subsection{Should {\em macroturbulent} broadening in massive stars be actually called {\em pulsational} broadening?}

Non-radial pulsations (NRP) as the shaping agent of the line-profiles of OB stars in addition to stellar rotation were suggested 
by \cite{Luc76} and \cite{Aer09} as a likely explanation for the {\em macroturbulent} broadening. 
In \cite{Sim10} we presented the 
first observational evidence of the existence of a tight connection between the size of this extra-broadening 
and parameters describing observed line-profile variations (LPVs) in a sample of 13 Sgs with spectral types ranging 
from O9.5 to B8. This result renders stellar oscillations the most probable physical origin of 
the extra-broadening in B Sgs. However, this is not the last word, some other tests are needed
to firmly declare NRP as the only physical phenomenon to explain the unknown broadening: (i) the same type 
of result should be reproduced for other stars in which the extra-broadening is
detected (including OB dwarfs, giants and supergiants); (ii) it must be observationally 
confirmed that the temporal behavior of the observed LPVs is compatible with the type of oscillations
considered by \cite{Aer09}; (iii) stars with a substantial {\em macroturbulent} broadening must be 
located inside the predicted instability strips in which stellar pulsations occur (or an even stronger
constraint, stars outside instability strips should not show any extra-broadening). 

\begin{figure}[t!]
\begin{minipage}[m]{8.cm}
\includegraphics[width=8.cm]{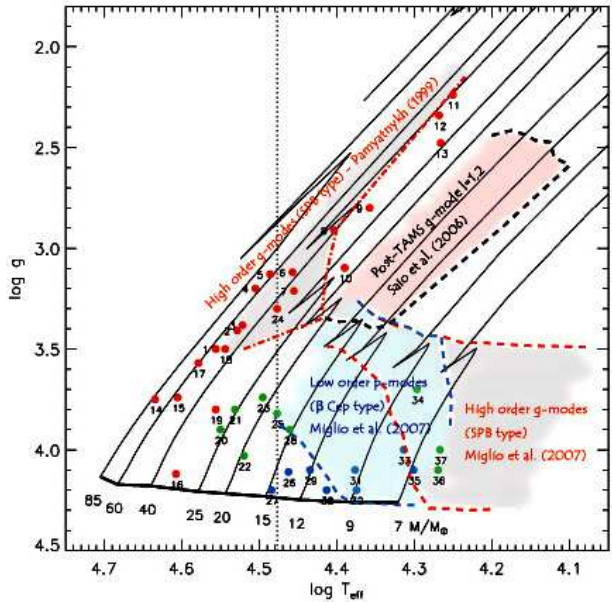}
\end{minipage}
\begin{minipage}[m]{6.cm}
\caption{\label{fig2} Location in the \grav\,--\,log\,\Teff\ diagram of a subsample of stars included 
in {\em IACOB} spectroscopic database, along with the instability domains 
for p- and g-mode oscillations predicted by \cite{Pam99}, \cite{Sai06}, and \cite{Mig07}. 
Three groups of stars with increasing size of the extra-broadening are
indicated in blue ($\le$\,20 \kms), green (20\,--\,40 \kms), and red ($\ge$\,40 \kms), 
respectively.}
\end{minipage}
\end{figure}

We are investigating these three points in parallel within the {\em IACOB} project. 
A preliminary analysis of time\,series spectra of four new targets 
(classified as O7.5\,Iabf, O9.5\,Iab, O6.5\,V((f))z, and B1\,I, respectively) obtained with the Mercator 
telescope during 8 nights seems to fit into the correlation found by \cite{Sim10}. 

The second point
requires long and extensive time\,series. In the last two years we have been compiling new high-resolution 
time\,series spectra for some of the stars studied in \cite{Sim10} (plus the four stars mentioned above) using 
the 1.2m Mercator and 2.5\,m NOT telescopes at the Roque de los Muchachos observatory (La Palma, Spain) and the 
2\,m NAO telescope at the Rozhen observatory (Bulgaria). The analysis of these observations is now in progress.

Regarding the last point, we are performing the quantitative spectroscopic analysis of the whole {\em IACOB} 
sample\footnote{Detected binary stars are excluded from this analysis} by applying the automatic tools 
presented in \cite{Sim11} and \cite{Cas12}. Figure \ref{fig2} indicates 
the location of $\sim$\,40 stars with spectral types ranging from O4 to B2 (including dwarfs, giants, 
bright giants and supergiants) in the \grav\,--\,log\,\Teff\ diagram along with the instability domains 
for p- and g-mode oscillations predicted by \cite{Pam99}, \cite{Sai06}, and \cite{Mig07}. The points 
corresponding to the analyzed stars are colored in blue, green and red as an indication of the size of 
the extra-broadening (in increasing order, respectively). 
It is important to remark that this figure represents a preliminary
overview of the work in progress\footnote{This refers mainly to the instability domains provided in the figure. 
While for stars with masses $\le$\,18\,$M_{\odot}$ we consider the new (revised) 
computations by \cite{Mig07} and \cite{Sai06}, for larger masses the plotted instability 
domain refers to the old computations by \cite{Pam99}. Although not included in the figure, we are 
aware of the recent works by \cite{Sai11} and M. Godart (PhD thesis, priv. comm.) which we 
plan to incorporate to our study.}.
However, it is already interesting to see the distribution of points
corresponding to stars with different characteristics (size) of the extra-broadening. In
particular, it is remarkable that while red and blue dots seems to support the pulsational origin (as
considered by \citeauthor{Aer09}, i.e., originated by low amplitude, high-order g modes),
the presence of green dots outside the instability domains motivates
the investigation of the possibility that other types of pulsations
could also mimic the {\em macroturbulent} profiles and OB dwarfs and giants. Indeed, low-amplitude
low-order p and g modes occur in stars in that part of the HRD and the
beating between such modes can also give rise to extra-broadening whenever the modes are not
properly resolved in a full pulsational analysis \citep[see, e.g.,][]{Mor06}.

\section{Macroturbulent broadening: a single snap-shot approach to investigate 
pulsations in massive stars?}\label{snapshot}

Observational asteroseismology is a very important field of stellar 
Astrophysics, but also very expensive in terms of observational time.
The resulting information extracted from the seismic 
data (and its modeling) of a given star is unique, since it is the only way to have 
access to the properties of the stellar interiors. However, very long and 
extensive time\,series of photometric and spectroscopic observations are 
needed to this aim. Obviously, there is no other way to be able to do a 
proper frequency analysis and mode identification, necessary for a seismic 
modeling; but, if the {\em macroturbulent} broadening -- NRP connection
is finally confirmed, this spectroscopic feature may perhaps also be used as 
a first, {\em less expensive} approach to investigate pulsations in massive stars. 
In particular, with just one spectrum of a given 
star (we need the spectrum to have a certain resolving power, but a very 
large signal-to-noise is not requested) we would be able to say if the
star is pulsating in the way expected from the fact that a {\em macroturbulent}
line shape is observed. We could hence analyze large samples of OB stars of
different spectral types and luminosity classes and easily conclude 
which of them have the ``same kind of" pulsational properties\footnote{A similar type 
of study was performed by \cite{Tel06} for the case of early B-type near-main
sequence stars without emission lines.} (roughly speaking!).
We could then correlate the size of the extra-broadening with other
stellar and wind properties. We could also select candidates to obtain the
spectroscopic and photometric time\,series necessary for a proper seismic
modeling, as in \cite{Sch04}. We may also use the type of result presented in Fig. \ref{fig2}
to obtain to a first order observational constraints for the predicted
instability domains for the type of pulsations where this spectroscopic feature 
is expected.
In view of this promising possibilities, the characterization and investigation of this
extra line-broadening in the context of stellar pulsations is more than justified. 



\end{document}